# Title: Electronic Structure of Superconducting Infinite-Layer Lanthanum Nickelates

**Authors**: Wenjie Sun[1,2]†, Zhicheng Jiang[3]†, Chengliang Xia[4]†, Bo Hao[1,2]†, Yueying Li[1,2,5], Shengjun Yan[1,2], Maosen Wang[1,2], Hongquan Liu[4,6], Jianyang Ding[7], Jiayu Liu[7], Zhengtai Liu[8], Jishan Liu[8], Hanghui Chen[4,9]*, Dawei Shen[3]*, Yuefeng Nie[1,2]*

**Affiliations:**

[1]National Laboratory of Solid State Microstructures, Jiangsu Key Laboratory of Artificial Functional Materials, College of Engineering and Applied Sciences, Nanjing University, Nanjing 210093, China.

[2]Collaborative Innovation Center of Advanced Microstructures, Nanjing University, Nanjing 210093, China.

[3]National Synchrotron Radiation Laboratory and School of Nuclear Science and Technology, University of Science and Technology of China, Hefei, 230026, China.

[4]NYU-ECNU Institute of Physics, NYU Shanghai, Shanghai, 200122, China.

[5]Department of Physics, Southern University of Science and Technology, Shenzhen, 518055, China.

[6]Department of Physics, Brown University, Providence, Rhode Island, 02912, USA.

[7]National Key Laboratory of Materials for Integrated Circuits, Shanghai Institute of Microsystem and Information Technology, Chinese Academy of Sciences, Shanghai, 200050, China.

[8]Shanghai Synchrotron Radiation Facility, Shanghai Advanced Research Institute, Chinese Academy of Sciences, Shanghai, 201210, China.

[9]Department of Physics, New York University, New York, New York, 10012, USA.

†These authors contributed equally to this work.

*Corresponding author. Email: hanghui.chen@nyu.edu (H.C.); dwshen@ustc.edu.cn (D.S.); ynie@nju.edu.cn (Y.N.)

**Abstract:**

Revealing the momentum-resolved electronic structure of infinite-layer nickelates is essential for understanding this new class of unconventional superconductors, but has been hindered by the formidable challenges in improving the sample quality. In this work, we report for the first time the angle-resolved photoemission spectroscopy of superconducting $La_{0.8}Sr_{0.2}NiO_2$ films prepared by molecular beam epitaxy and *in situ* atomic-hydrogen reduction. The measured Fermi topology closely matches theoretical calculations, showing a large Ni-$d_{x^2-y^2}$ derived Fermi sheet that evolves from hole-like to electron-like along $k_z$, and a three-dimensional (3D) electron pocket centered at Brillouin zone corner. The Ni-$d_{x^2-y^2}$ derived bands show a mass enhancement ($m^*/m_{DFT}$) of 2-3, while the 3D electron band shows negligible band renormalization. Moreover, the Ni-$d_{x^2-y^2}$ derived states also display a band dispersion anomaly at higher binding energy, reminiscent of the waterfall feature and kinks observed in cuprates.

**Main text:**

Square-planar nickelates with $Ni^{1+}$ valence states have attracted considerable attention since the seminal discovery of superconductivity in hole-doped infinite-layer $R$NiO$_2$ ($R$ = La, Pr, Nd) (*1-3*). Mimicking $3d^9$ electron configuration and layered lattice structure as high-$T_c$ cuprates, infinite-layer nickelates serve to be an ideal new platform for testing various theoretical models and uncovering the mechanisms behind unconventional high-$T_c$ superconductivity. Despite sharing many similarities with cuprates, important differences in electronic structures and correlation effects have also been observed or proposed in nickelates. For instance, the multiband features especially the presence of three-dimensional (3D) conduction bands near the Fermi level have been pointed out both experimentally (*4, 5*) and theoretically (*6-8*), leading to self-doping effects, which may explain the metallic ground state (*1, 9, 10*) and the absence of long-range antiferromagnetic orders (*11-14*). In addition, while cuprates are in the charge-transfer regime according to the Zaanen-Sawatzky-Allen (ZSA) scheme (*15*), infinite-layer nickelates are closer to the Mott-Hubbard regime because their charge transfer gap between Ni-$3d$ and O-$2p$ orbitals is large (*4, 16, 17*). Moreover, the superconducting pairing symmetry in nickelates is still under hot debate (*18-25*), compared to the widely accepted $d$-wave symmetry in cuprates. In the quest to clarify these debates, it is imperative to directly reveal the momentum-resolved electronic structure for infinite-layer nickelates, which remains lacking to date.

The major obstacle is the lack of a high-quality sample surface, which is a prerequisite for surface-sensitive electronic structure characterizations like angle-resolved photoemission spectroscopy (ARPES). Typically, superconducting infinite-layer nickelates are synthesized by a post-reduction process, during which the apical oxygens in the perovskite precursor phase are removed using metal hydrides (e.g., CaH$_2$) and the surface is known to degrade unavoidably. Fortunately, recent advances in molecular beam epitaxy (MBE) growth of nickelate films (*26-28*) and *in situ* atomic-hydrogen reduction allow for the preparation of superconducting nickelates with atomically smooth surface *in vacuo* (*29, 30*), providing suitable samples with high surface quality for ARPES measurements.

In this work, we study the electronic structure of superconducting La$_{0.8}$Sr$_{0.2}$NiO$_2$ thin films by synchrotron-based ARPES and theoretical many-body calculations. The thin film samples were prepared by using the combination of oxide MBE and *in situ* atomic-hydrogen reduction (Fig. S1). Our results reveal a large Fermi surface sheet that is derived from Ni-$d_{x^2-y^2}$ orbital, which shows a mass enhancement of 2-3, similar to that in cuprates (*31, 32*). In addition, a small 3D electron pocket derived from the conduction bands appears at the Brillouin zone (BZ) corner, which barely shows mass enhancement. Moreover, high energy dispersion anomalies are observed for the Ni-$d_{x^2-y^2}$ derived pockets, which resemble the waterfall features and kinks in cuprates.

**Fermi surface of La$_{0.8}$Sr$_{0.2}$NiO$_2$**

Superconducting La$_{0.8}$Sr$_{0.2}$NiO$_2$ thin films were prepared on TiO$_2$-terminated SrTiO$_3$ substrates, exhibiting superconducting onset and zero-resistance temperature of 14.1 and 11.9 K, respectively (Fig. 1B, and see Methods for details of growth and reduction processes), and the surface quality has been confirmed by scanning transmission electron microscopy as reported previously (*30*). The ARPES measurements were carried out at 6 K using 72- and 106-eV photons, corresponding to $k_z = \pi$ and $k_z = 0$ plane, respectively, as determined from the photon-energy-dependent measurements (Fig. S2).

In general, the Fermi surface topology in our ARPES measurements agrees well with the theoretical calculations. Figure 1A shows the calculated 3D Fermi surface of La$_{0.8}$Sr$_{0.2}$NiO$_2$ using virtual crystal approximation treatment without considering the on-site Coulomb interaction ($U = 0$ eV) (see Methods for calculation details). Besides the major Ni-$d_{x^2-y^2}$ derived sheets extending across the whole range of $k_z$, there are additional electron pockets with the hybridized character of Sr/La $d_{xy}$ and Ni $d_{xz}/d_{yz}$ orbitals at the zone corners (Fig. S3). The experimental Fermi surface maps taken at different $k_z$ are shown in Fig. 1, C and D, which were measured at the second BZ to get an improved signal-to-noise ratio, as also observed in Ruddlesden–Popper (RP)-type nickelates (*33-35*). At the $k_z = 0$ plane, the overall feature is akin to the cuprates, with large Ni-$d_{x^2-y^2}$ derived hole pockets centered at $M$ ($\pi,\pi,0$) point (Fig. 1C). However, distinct from the nearly $k_z$-independent dispersions in optimally-doped cuprates, the Ni-$d_{x^2-y^2}$ derived pockets display obvious $k_z$ dispersions in nickelates, turning from hole-like at the BZ center to electron-like at the BZ top/bottom. In addition, small electron pockets are also resolved at the zone corners at the $k_z = \pi$ plane (Fig. 1D). In Fig. 1, E and F, Fermi momenta are extracted from the observed Fermi surface maps, where both pockets match the calculations. Note that no electron pocket is observed at the zone center, aligning with theoretical prediction that hole doping depletes such electron pocket in the parent compound (*7*).

**Band renormalization of the Ni-$d_{x^2-y^2}$ derived bands**

Figure 2 shows representative energy-momentum spectra of the Ni-$d_{x^2-y^2}$ derived bands along different cuts. Consistent with the density functional theory (DFT) calculations (Fig. S3), the Ni-$d_{x^2-y^2}$ orbital character is also confirmed by our polarization-dependent measurements (Fig. S4). From the observed band dispersions, one can see nearly momentum-independent feature at higher binding energy, which is known as the dispersion anomaly (*34-42*) and will be discussed later in detail. Above this dispersion anomaly, the bands near the Fermi level exhibit clear momentum dependence, showing visible band renormalization effects compared to the calculated band dispersions (blue solid lines in Fig. 2, G to I). The renormalization factors are determined by comparing the measured dispersions that are extracted from peak positions in the corresponding momentum distribution curves (MDCs) with scaled calculated bands (blue dashed lines in Fig. 2, G to I). The obtained mass enhancement factors ($m^*/m_{DFT}$, where $m_{DFT}$ is the effective mass derived from the DFT bands) are around 2-3 for different momentum cuts, which are very close to that observed in the normal state of cuprates (*31, 32*). Interestingly, this mass enhancement factor is also comparable to that of trilayer (La$_4$Ni$_3$O$_{10}$) (*33*) and 2222-type double-layer (*34*) nickelates, but is smaller than that observed in reduced trilayer (Pr$_4$Ni$_3$O$_8$) (*43*) and 1313-type double-layer (*35*) nickelates. The reason behind such commonality/discrepancy regarding the mass enhancement factor of Ni-$d_{x^2-y^2}$ orbital in nickelates may lie in the electronic correlation strength, crystal structure, and carrier doping level, which warrants further investigation. Moreover, the quasiparticle mean free path is known to be inversely proportional to the full width at half maximum (FWHM) of the MDC peaks (*44*), and is extracted to be around 8 Å for Ni-$d_{x^2-y^2}$ derived bands (Fig. S5), which is consistent with the fact that superconductivity in nickelates exists in the dirty limit (*26, 45*).

In order to gain more insights into the many-body electronic structure of infinite-layer nickelates, we also construct a low-energy effective model to take into account local Coulomb interaction on Ni $d$ orbitals and use dynamical mean field theory (DMFT) to calculate the many-body spectral function $A(\mathbf{k}, \omega)$ of La$_{0.8}$Sr$_{0.2}$NiO$_2$ (see Methods for details). The non-interacting part of the low-energy effective model is obtained by downfolding the DFT band structure (see Fig. S6). The calculated many-body spectral function $A(\mathbf{k}, \omega)$ and the experimental ARPES data are compared in Figs. 3 as well as in S7. We find a good agreement between theory and experiment close to the Fermi level if we take $U_d$ = 3.9 eV in the calculation. In our low-energy effective model, $U_d$ = 3.9 eV yields an effective mass of 3.0 $m_{DFT}$ for Ni-$d_{x^2-y^2}$ orbital (see Fig. S8 for details).

**Dispersion anomaly in the Ni-$d_{x^2-y^2}$ derived bands**

Apart from the renormalized dispersions close to the Fermi energy, prominent electronic dispersion anomaly is observed in Ni-$d_{x^2-y^2}$ derived bands at various momentum cuts, exhibiting a kink at 70-100 meV and extending up to 300 meV and beyond (Figs. 2 and 3). This phenomenon bears a resemblance to the high-energy waterfall dispersion observed in cuprates (*36-41*), as well as in RP-type nickelates (*34, 35, 42*). In cuprates, the characteristic energy scale of the waterfall dispersion is 300-500 meV, and several scenarios have been discussed, including the effects of interactions by bosonic modes (*36*), spin charge separation (*38*), spin fluctuations (*40, 46*), and photoemission matrix element effects (*39, 41*). Unlike the waterfall dispersion in cuprates that starts at hundreds of meV, however, the dispersion anomaly in infinite-layer nickelates looms at a lower energy, around 70-100 meV, a value close to the low-energy kink observed in cuprates. In cuprates, the low-energy kink at about 70 meV has been ascribed to the electron coupling to a bosonic mode (phonon (*31*) or spin excitation (*47*)). Indeed, our calculated phonon dispersions for undoped LaNiO$_2$ also show a nearly dispersion-less branch of optical phonon located near 65 meV, with a similar in-plane full-breathing stretching mode as that in optimally-doped cuprates (Fig. S9). Note that Sr doping will elevate this phonon frequency due to the lighter atomic mass of Sr than La. Hence, the approximate alignment of the onset energy of the kink with this optical phonon mode suggests a potential role of electron-phonon coupling in the emergence of unconventional superconductivity in infinite-layer nickelates, although the strength of such kinks varies within different theoretical models (*48, 49*). It is noted that the magnetic excitations (*11*) and spin fluctuations (*14*) also appear in doped infinite-layer nickelates, which may contribute to such dispersion anomaly as well. Nonetheless, the exact role of electron-phonon coupling and what other types of bosons with different energy scales are also involved in shaping such a dispersion anomaly in infinite-layer nickelates remain important open questions for further exploration.

**Band dispersion of the three-dimensional electron bands**

As a major difference between cuprates and nickelates, the electron bands are also believed to play a certain role in nickelates (*20, 21*). As aforementioned, we only observe the electron pockets at the zone corners, consistent with the theoretical prediction that the hole doping suppresses the electron pocket at the zone center. Compared to the $d_{x^2-y^2}$-derived bands, the conduction bands disperse down to ~300 meV without any dispersion anomaly, as shown in Fig. 4. More importantly, it barely shows renormalization effect, indicating negligible electron correlations of this electron band. This result can be explained by different orbital characters and filling for the bands, that is, the partially-filled conduction bands are derived from a mixture of Sr/La $d_{xy}$ and Ni $d_{xz}/d_{yz}$ orbitals (*7*), both of which exhibit weaker electron correlations compared to the half-filled $d_{x^2-y^2}$ derived bands. Note that an additional electron pocket with much weaker intensity and shallower dispersion (band bottom at ~100 meV) is also vaguely resolved (Fig. 4B), but not captured by our theoretical calculations. Due to the enormous challenge of preparing such samples, there may be trace amount of regions containing residual apical oxygens, which could result in hole doping effects and elevate the bands toward the Fermi level. Currently, we cannot rule out other possibilities associating with this additional feature, including interactions between electrons and other quasiparticles, such as plasmons (*50, 51*).

**Discussion and conclusions**

It is worth noting that the ARPES spectra shown above are all obtained with relatively high photon energies. At these photon energies, the best energy resolution is around 15 meV, which is too large to resolve the superconducting gap of a few meVs in infinite-layer nickelates as determined by scanning tunnelling spectroscopy (*21*), terahertz spectroscopy (*22, 23*), and London penetration depth measurements (*24, 25*). To improve the energy resolution, we also performed measurements with lower photon energies but the electronic states near the Fermi level became too weak to be measured (not shown), which might be related to the strong matrix element effects. As a result, no conclusive results can be drawn on the superconducting gap size and symmetry in the present study so far. Improving the ARPES resolution as well as the sample quality may shed light on the superconducting gap size and symmetry and will be explored in the future.

In summary, the momentum-resolved electronic structure of infinite layer nickelates has been revealed by performing ARPES measurements on high-quality films prepared using a combination of MBE and *in situ* atomic hydrogen reduction. The overall Fermi topology agrees well between experimental observations and theoretical calculations. The Ni-$d_{x^2-y^2}$ derived bands exhibit a mass enhancement of 2-3 and a giant dispersion anomaly, similar to high-$T_c$ cuprates. In contrast, the 3D electron bands derived from a mixture of Sr/La $d_{xy}$ and Ni $d_{xz}/d_{yz}$ orbitals show negligible band renormalization and exhibit no kinks and dispersion anomaly. Our work presents the first direct experimental observation of momentum-resolved electronic structures in the superconducting infinite-layer nickelates, providing essential guidelines for the exploration of unconventional superconductivity in these materials.

**Acknowledgements:**

We acknowledge insightful discussions with Harold Y. Hwang. H.C. acknowledge the High-Performance-Computing of New York University (NYU-HPC) for providing the computational resources. **Funding:** This work was supported by the National Key R&D Program of China (Grant Nos. 2021YFA1400400, 2022YFA1402502, 2023YFA1406304, and 2021YFE0107900); National Natural Science Foundation of China (Grant Nos. 11861161004, U2032208, and 12374064) and the Fundamental Research Funds for the Central Universities (Grant No. 0213−14380221). H.C. is supported by the Science and Technology Commission of Shanghai Municipality under grant number 23ZR1445400 and a grant from the New York University Research Catalyst Prize. Part of this research used Beamline 03U of the Shanghai Synchrotron Radiation Facility, which is supported by $ME^2$ project under Contract No.11227902 from National Natural Science Foundation of China. **Author contributions:** Y.N. conceived the project and designed the ARPES experiments with D.S. W.S., Z.J., B.H., S.Y., M.W., J.D., J.L., Z.L., and J.L. conducted the experiment at Shanghai Synchrotron Radiation Facility. C.X., H.L., and H.C. developed theoretical models and performed calculations. W.S., B.H., Y.L., S.Y., and M.W. synthesized and characterized nickelate films for the experiment., H.C., D.S., and Y.N. discussed and interpreted the results. W.S., B.H., H.C., and Y.N. wrote the manuscript with input from all authors. **Competing interests:** The authors declare no conflict of interest. **Data and materials availability:** The data that support the findings of this study are available from the corresponding author on reasonable request.

**List of Supplementary Materials:**

Materials and Methods
Figs. S1 to S9
Table S1

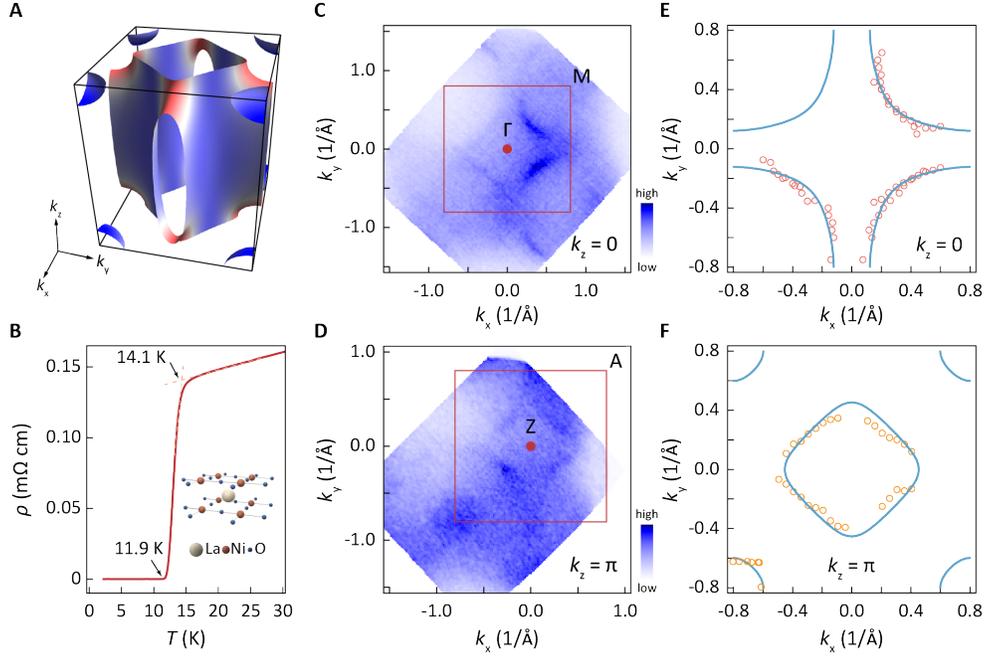

**Fig. 1 Fermi surface of $La_{0.8}Sr_{0.2}NiO_2$.**

(**A**) DFT-calculated Fermi surface of $La_{0.8}Sr_{0.2}NiO_2$, consisting of Ni-$d_{x^2-y^2}$ derived pockets and three-dimensional electron pockets at Brillouin zone (BZ) corners. (**B**) Temperature-dependent resistivity curve at low temperature, showing characteristic $T_{c,onset}$ and $T_{c,zero}$ of 14.1 and 11.9 K, respectively. Inset shows the crystal structure of the infinite-layer structure. (**C** to **D**) Unsymmetrized Fermi surface maps taken at $k_z = 0$ and $k_z = \pi$ plane in the second BZ, with photon energy of 106 and 72 eV, respectively. The photoemission intensity is integrated over ± 15 meV with respect to the Fermi level. (**E** to **F**) Fermi momenta ($k_F$) extracted from corresponding momentum distribution curves (MDCs) of Fermi surface maps in (C), and second derivatives of that in (D), respectively (marked by hollow circles). The DFT-calculated Fermi surface (solid lines) is overlaid for comparison. Error bars are within the corresponding symbols.

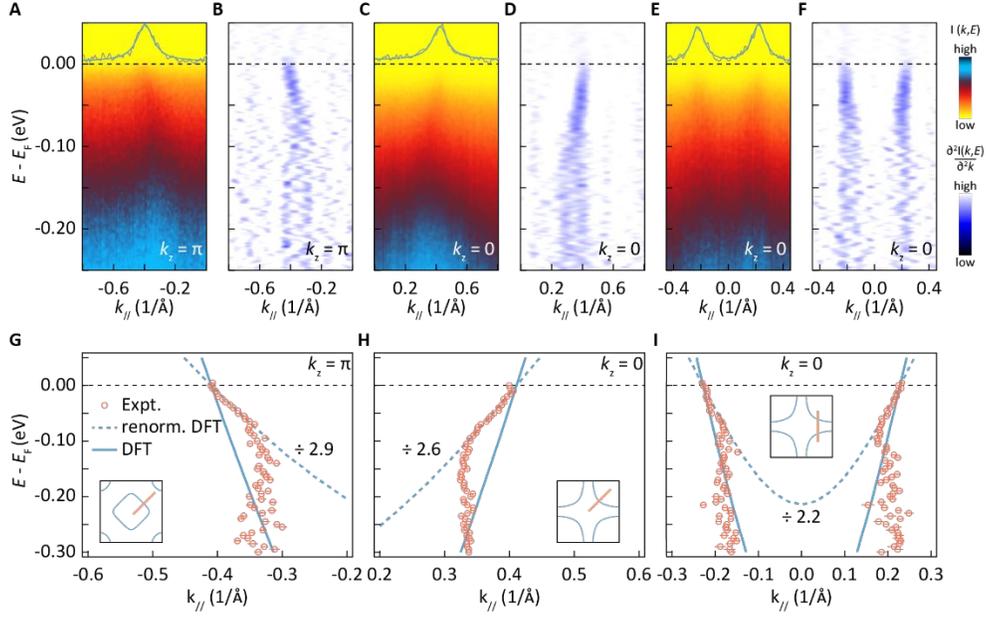

**Fig. 2 Low-energy band renormalization and high-energy dispersion anomaly of Ni-$d_{x^2-y^2}$ derived bands.**

(**A** to **B**) Measured ARPES spectrum and corresponding second-derivative spectrum taken along orange solid line in the inset in (G) at $k_z = \pi$ plane. Note that the cut is slightly off the diagonal direction of the BZ. (**C** to **D**) Spectrum taken along orange solid line in the inset in (H) at $k_z = 0$ plane. (**E** to **F**) Spectrum taken along orange solid line in the inset in (I) at $k_z = 0$ plane. Insets in (A), (C), and (E) are the MDCs (gray lines) at the Fermi energy, which are fitted by Lorentzian peaks (blue lines) after subtracting a linear background. (**G** to **I**) Band dispersions extracted from the peak positions of MDCs in (A), (C), and (E), respectively. The original DFT-calculated bands (blue solid lines) are scaled by a renormalization factor shown in each panel to match the experimental dispersions. And only dispersions above the binding energy of 100 meV are compared with the scaled DFT results, below which the dispersion anomaly appears, as discussed in the main text. All spectra are measured at 6 K unless otherwise mentioned.

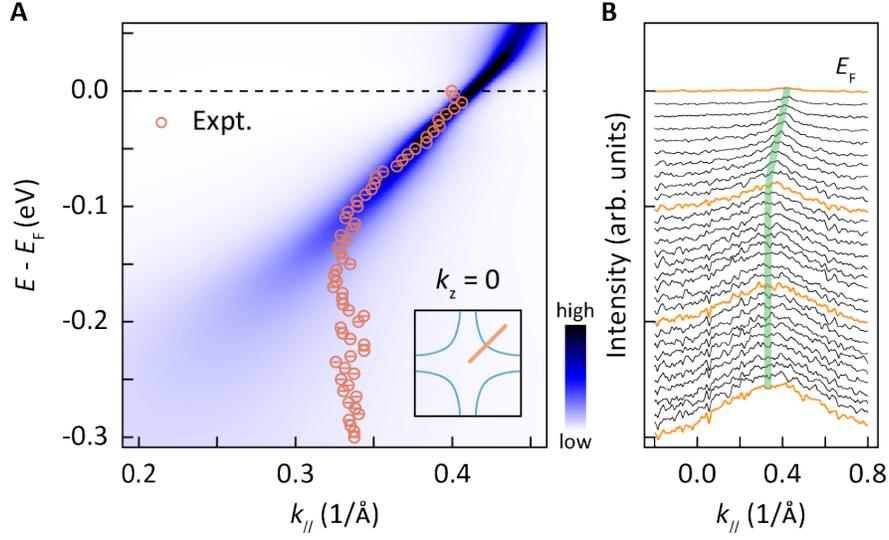

**Fig. 3 DMFT spectral function for Ni-$d_{x^2-y^2}$ derived bands.**

(**A**) Corresponding dynamical mean field theory (DMFT) spectral function for the measured ARPES spectrum shown in Fig. 2C. The measured dispersion can be well reproduced when $U_d$ = 3.9 eV, giving a mass enhancement factor of 3.0. (**B**) Corresponding waterfall plot of MDCs ranging from -0.3 eV to the Fermi level. Shadow green line illustrates the band dispersion and is the guide to the eyes.

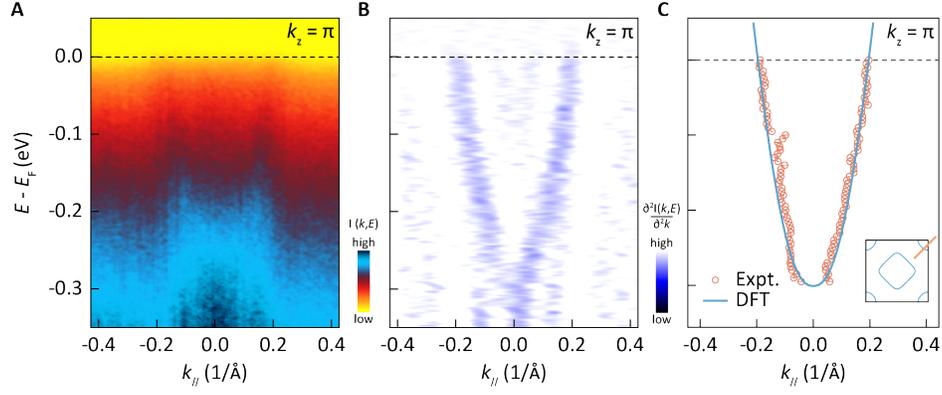

**Fig. 4 Electron pockets around BZ corners.**

(**A**) Measured ARPES spectrum taken along the orange solid line in the inset in (C) at $k_z = \pi$ plane. Note that the cut is slightly off the diagonal direction of the BZ. (**B**) Corresponding second-derivative spectrum of (A). (**C**) Peak positions extracted from MDCs in (A). Blue solid line corresponds to the DFT-calculated results. The calculated band dispersion shows excellent agreement with the experimental data, suggesting negligible electron correlations for the electron bands.

# Supplementary Materials for

# Electronic Structure of Superconducting Infinite-Layer Lanthanum Nickelates

Wenjie Sun[1,2]†, Zhicheng Jiang[3]†, Chengliang Xia[4]†, Bo Hao[1,2]†, Yueying Li[1,2,5], Shengjun Yan[1,2], Maosen Wang[1,2], Hongquan Liu[4,6], Jianyang Ding[7], Jiayu Liu[7], Zhengtai Liu[8], Jishan Liu[8], Hanghui Chen[4,9]*, Dawei Shen[3]*, Yuefeng Nie[1,2]*

*Correspondence to: Hanghui Chen, hanghui.chen@nyu.edu, Dawei Shen, dwshen@ustc.edu.cn, Yuefeng Nie, ynie@nju.edu.cn

**The PDF file includes:**

    Materials and Methods
    Figs. S1 to S9
    Table S1

## Materials and Methods

### Thin film growth and *in situ* reduction

The perovskite $La_{0.8}Sr_{0.2}NiO_3$ thin films were grown on (001)-orientated $TiO_2$-terminated $SrTiO_3$ substrates using reactive molecular-beam epitaxy in a DCA R450 system. Growth was performed at substrate temperatures around 600 °C and under an oxidant (distilled ozone) background partial pressure of $1\times10^{-5}$ torr. Rough fluxes of the La, Ni and Sr sources were calibrated using a quartz crystal microbalance. And the precise fluxes were determined based on the reflection high-energy electron diffraction (RHEED) oscillations, when growing $LaNiO_3$ and $SrTiO_3$ films on $SrTiO_3$ substrates using the co-deposition method (*27*). After growth, the perovskite films were cooled to 200 °C under the same background pressure, and then *in situ* transferred into an UHV reduction chamber with base pressure better than $1\times10^{-9}$ torr. The atomic hydrogen was generated by a home-modified e-beam source (Single Pocket Electron Beam Evaporator, SPECS) and irradiated films in the direction normal to the surface. The reduction parameters were optimized as reported previously (*30*).

### ARPES measurements

After *in situ* reduction, the films were transferred to beamlines BL03U (*52*) and BL09U at the Shanghai Synchrotron Radiation Facility (SSRF), using a vacuum suitcase with base pressure better than $1\times10^{-10}$ torr. Data shown in the main text were measured at BL03U with the base pressure of $5\times10^{-11}$ Torr and a DA30 analyzer (Scienta-Omicron) at 6 K. The linear horizontal polarized light orthogonal to the analyzer slit was utilized during measurements. The highest energy resolution was 15 meV with photon energy of 72 eV in our experiments, and the angular resolution was set to be 0.2°. The Fermi energy is determined by a polycrystalline gold which is electrically connected to the film. Polarization-dependent and parts of the photon-energy dependent measurements were performed at BL09U with the base pressure of $5\times10^{-11}$ Torr and a DA30 analyzer (Scienta-Omicron) at 14 K.

### Structural, topography and transport characterization

X-ray diffraction measurements were performed using a Bruker D8 Discover diffractometer with a monochromated Cu-$K_\alpha$ ($\lambda$ = 1.5418 Å) radiation. The surface topography was characterized by an Asylum Research MFP-3D Origin+ scanning probe microscope using the contact mode. The electrical transport measurements were performed either at a Teslatron PT system (Oxford

Instruments) or a physical property measurement system (Quantum Design) using the standard van de Pauw geometry. The ohmic contacts were achieved by ultrasonic wire bonding using aluminum wires.

**Band structure calculations**

We first perform density functional theory (DFT) calculations (*53, 54*). The DFT method is implemented in the Vienna ab initio simulation package (VASP) code (*55*) with the projector augmented wave (PAW) method (*56*). The Perdew-Burke-Ernzerhof (PBE) (*57*) functional is used as the exchange-correlation functional in DFT calculations. The empty La-4$f$ orbitals are included in the pseudopotential. To treat with the Sr doping, we use the virtual crystal approximation (*58*). We use an energy cutoff of 600 eV and sample the Brillouin zone by using $\Gamma$-centered $k$-mesh of 14 × 14 × 14 per primitive cell. To simulate the experimental setup of La$_{0.8}$Sr$_{0.2}$NiO$_2$ thin films grown on SrTiO$_3$ substrates, we impose the constraint that the in-plane lattice constants ($a$ and $b$) are fixed to be the theoretical lattice constant of SrTiO$_3$ (3.948 Å, which is 1.1% larger than the experimental value). The $c$-axis and internal atomic coordinates are fully relaxed with an energy convergence criterion of $10^{-7}$ eV, force convergence criterion of 0.001 eV/Å, and strain convergence of 0.1 kbar. The DFT-optimized $c$-axis lattice constant of La$_{0.8}$Sr$_{0.2}$NiO$_2$ is 3.369 Å.

**Low-energy effective model**

To study the spectral function around the Fermi level, we perform a downfolding procedure and construct a low-energy effective model, following our previous study (*59*). The effective model, known as the *ds* model, consists of 2 orbitals: a Ni-$d_{x^2-y^2}$ orbital at (0.5, 0.5, 0.5) and an effective-$s$ orbital at (0.5, 0.5, 0.0). The model can be compactly written as:

$$\hat{H} = \sum_{\mathbf{k}\sigma} \hat{\Psi}^{\dagger}_{\mathbf{k}\sigma} \mathcal{H}_0(\mathbf{k}) \hat{\Psi}_{\mathbf{k}\sigma} + U_d \sum_i \hat{n}^d_{i\uparrow} \hat{n}^d_{i\downarrow} - \hat{V}^{\mathrm{dc}} \qquad (1)$$

where $\hat{\Psi}^{\dagger}_{\mathbf{k}\sigma} = (\hat{d}^{\dagger}_{\mathbf{k}\sigma}, \hat{s}^{\dagger}_{\mathbf{k}\sigma})$ are the creation operators on Ni-$d_{x^2-y^2}$ and effective-$s$ orbitals with momentum **k** and spin $\sigma$. $\hat{n}^d_{i\sigma} = \hat{d}^{\dagger}_{i\sigma} \hat{d}_{i\sigma}$ is the occupancy operator of Ni-$d_{x^2-y^2}$ orbital at site $i$ with spin $\sigma$. $\mathcal{H}_0(\mathbf{k})$ is a 2 × 2 matrix:

$$\mathcal{H}_0(\mathbf{k}) = \begin{bmatrix} \epsilon_d(\mathbf{k}) & V_{ds}(\mathbf{k}) \\ V^*_{ds}(\mathbf{k}) & \epsilon_s(\mathbf{k}) \end{bmatrix} \qquad (2)$$

The energy dispersion and hybridization terms are:

$$\epsilon_d(\mathbf{k}) = 2t_d^{100}\left(\cos k_x + \cos k_y\right) + 4t_d^{110}\cos k_x \cos k_y + 2t_d^{200}\left(\cos 2k_x + \cos 2k_y\right) \quad (3)$$

$$\begin{aligned}\epsilon_s(\mathbf{k}) = E_{ds} &+ 2t_s^{100}\left(\cos k_x + \cos k_y\right) + 2t_s^{001}\cos k_z + 4t_s^{110}\cos k_x \cos k_y \\ &+ 4t_s^{101}\left(\cos k_x + \cos k_y\right)\cos k_z + 8t_s^{111}\cos k_x \cos k_y \cos k_z\end{aligned} \quad (4)$$

$$V_{ds}(\mathbf{k}) = 2t_{ds}^{100}\left(\cos k_x - \cos k_y\right)(1 + e^{-ik_z}) + 2t_{ds}^{200}\left(\cos 2k_x - \cos 2k_y\right)(1 + e^{-ik_z}) \quad (5)$$

The onsite energy difference $E_{ds}$ and hopping parameters are obtained by fitting to the near-Fermi-level DFT band structure of La$_{0.8}$Sr$_{0.2}$NiO$_2$ (Fig. S6). The fitted values are shown in Table S1.

**DMFT calculations**

Then we perform dynamical mean field theory (DMFT) calculations (*60, 61*) to solve the *ds* model Eq. (1). We employ the continuous-time quantum Monte Carlo (CTQMC) algorithm based on the hybridization expansion (*62, 63*). The impurity solver is developed by K. Haule (*64*). For each DMFT iteration, a total of 1 billion Monte Carlo samples are collected to converge the impurity Green's function and self-energy. We set the temperature to be 116 K. We check all the main results at a lower temperature of 58 K and no significant difference is found. We use the fully-localized-limit (FLL) double counting, which reads:

$$V^{dc} = U_d\left(N_d - \frac{1}{2}\right) \quad (6)$$

To obtain the spectral functions, the imaginary axis self energy is continued to the real axis by using the maximum entropy method (*65*). Then the real axis momentum-dependent Green's function is calculated by using the Dyson equation.

**Self-energy and the Ni effective mass from the *ds* model**

In this section, we study the effective mass of Ni-$d_{x^2-y^2}$ orbital in the low-energy effective model (Eq. (1)). Since the state is metallic, in local DMFT calculations, we can directly relate the effective mass to the quasi-particle weight Z which is defined from the real part of the retarded self-energy on the real-frequency axis:

$$\frac{m^*}{m_{DFT}} = \frac{1}{Z} = 1 - \frac{\partial \text{Re}\Sigma(\omega + i0^+)}{\partial \omega}\bigg|_{\omega=0} \quad (7)$$

However, the self-energy from CTQMC is calculated on the Matsubara frequencies. If the low-frequency properties are reasonably well described by the Fermi-liquid fixed point, the low-frequency limit of the real-frequency self-energy may be inferred with reasonable accuracy from the data at small Matsubara frequencies (66):

$$\frac{m^*}{m_{DFT}} = \frac{1}{Z} \simeq 1 - \frac{d\mathrm{Im}\Sigma(i\omega_n)}{d\omega_n}\bigg|_{\omega_n \to 0} \quad (8)$$

In practice, we extract the effective mass by fitting a fourth-order polynomial to the first six Matsubara-axis data points for $\mathrm{Im}\Sigma(i\omega_n)$ and computing the needed quantities from the fitting function. Fig. S8 shows $\mathrm{Im}\Sigma(i\omega_n)$ calculated by using the low-energy effective model at $U_d = 3.9$ eV. Panel (a) shows a large frequency range and panel (b) shows the self-energy of the first six Matsubara frequencies. From the fitting, we obtain that at $U_d = 3.9$ eV, the effective mass of Ni-$d_{x^2-y^2}$ orbital ($m^*/m_{DFT}$) is 3.0.

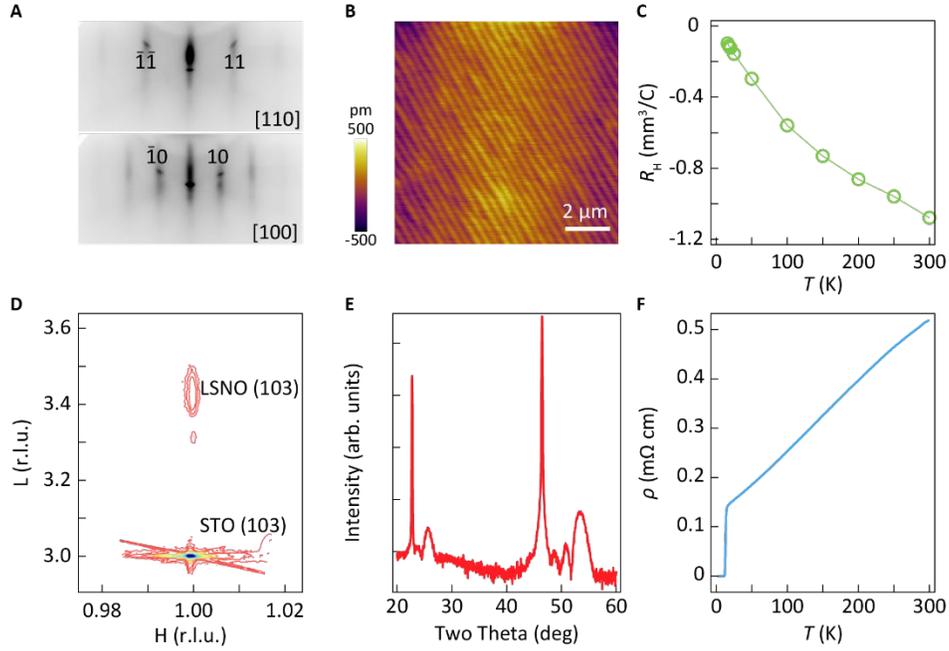

**Fig. S1.**

**Additional structural and transport characterizations of a 20-u.c.-thick La$_{0.8}$Sr$_{0.2}$NiO$_2$ thin film.** **(A)** Reflection high-energy electron diffraction (RHEED) patterns of as-grown La$_{0.8}$Sr$_{0.2}$NiO$_3$ thin film taken along [110] and [100] directions. **(B)** Atomic force microscopy of *in situ* reduced La$_{0.8}$Sr$_{0.2}$NiO$_2$ film, revealing atomically smooth surface. **(C)** Temperature-dependent Hall coefficient of La$_{0.8}$Sr$_{0.2}$NiO$_2$ film. **(D)** Reciprocal space map around (103) diffraction spot of SrTiO$_3$ (STO) substrate, demonstrating that the La$_{0.8}$Sr$_{0.2}$NiO$_2$ (LSNO) film is fully in-plane strained to the STO substrate. **(E)** 2$\theta$-$\omega$ x-ray diffraction scan of La$_{0.8}$Sr$_{0.2}$NiO$_2$ film. **(F)** Same temperature-dependent resistivity curve as shown in Fig. 1B in the main text, but with larger temperature range (2-300 K).

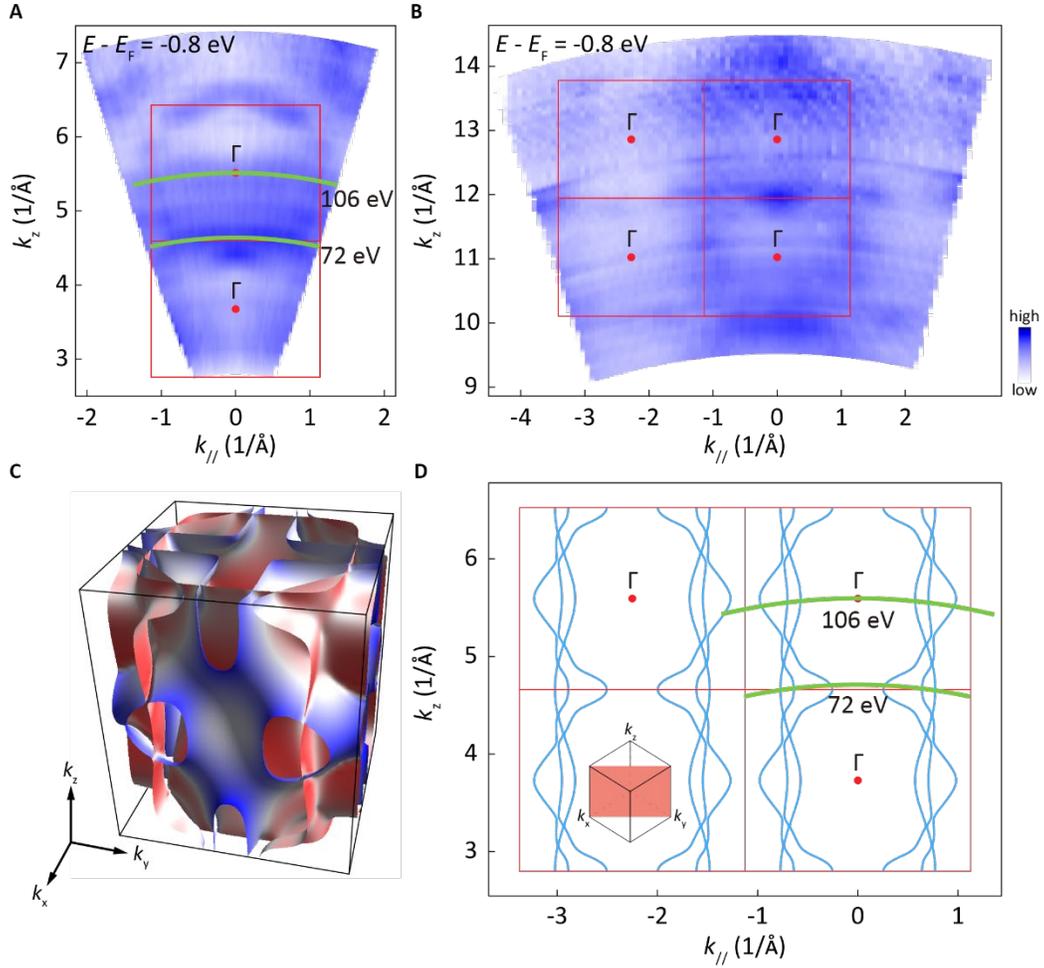

**Fig. S2.**
**$k_z$ dispersions of La$_{0.8}$Sr$_{0.2}$NiO$_2$ thin film. (A-B)** $k_z$ dispersions at binding energy of 0.8 eV taken along the diagonal plane of the three-dimensional BZ in (C), using photon energies between 20-200 and 335-794 eV, respectively. The inner potential was set to 15 eV. According to the observed dispersions, 72- and 106-eV photons are near $k_z = \pi$ and $k_z = 0$ plane, respectively. Note that the spectral weight of the electronic states near the Fermi level is too weak to show clear feature. **(C)** Calculated constant-energy surface of three Ni $t_{2g}$ bands ($d_{xy}$, $d_{xz}$, and $d_{yz}$). **(D)** Calculated constant-energy maps taken along the diagonal plane of three-dimensional BZ in (C), shown as the red shaded plane in the inset. Red boxes represent BZs. The overall features of the observed $k_z$ dispersions are consistent with the calculations.

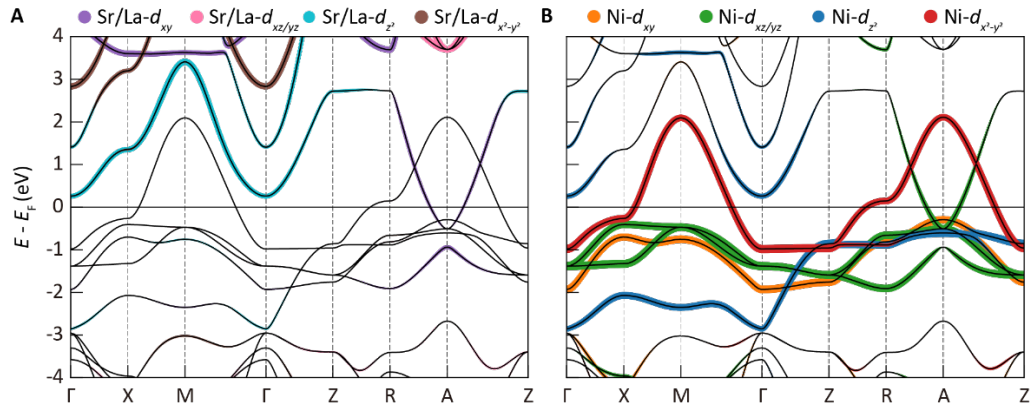

**Fig. S3.**
**DFT-calculated band structure of La$_{0.8}$Sr$_{0.2}$NiO$_2$ along high symmetric k-path. (A-B)** The weightings of different orbital character for Sr/La and Ni are projected onto the DFT-calculated band structure, respectively. Bands at $\Gamma$ point show mainly Ni-$d_{x^2-y^2}$ character, while they show the mixture of Sr/La-$d_{xy}$ and Ni-$d_{xz}/d_{yz}$ character at A point.

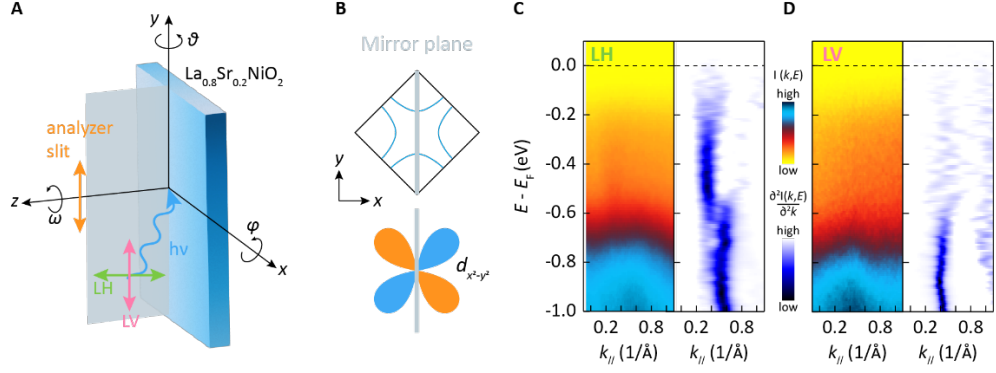

**Fig. S4.**
**Polarization-dependent ARPES measurements.** **(A)** Experimental geometry for ARPES measurements. During measurements, $La_{0.8}Sr_{0.2}NiO_2$ films are rotated along $x$, $y$, and $z$ axis to make sure the photoemission signal coming from the second Brillouin zone (BZ). Therefore, the incident angle of the synchrotron radiation with respect to the film surface varies with the photon energies. The slit of the analyzer is along the y axis. Linear horizontal (LH) and linear vertical (LV) polarized light is used to distinguish orbital characters. **(B)** Schematic drawing of the experimental geometry highlighting the matrix element effects for the Ni-$d_{x^2-y^2}$ orbital. The film diagonal direction is aligned along the analyzer slit ($\omega = 45°$). In such configuration, the parity of LH and LV is odd and even respectively, while the parity of Ni-$d_{x^2-y^2}$ orbital is odd. Therefore, the stronger photoemission intensity from Ni-$d_{x^2-y^2}$ orbital is expected for LH polarized light. **(C-D)** Spectra and corresponding second-derivatives taken along gray solid line in (B) with LH and LV polarized light, respectively, suggesting major Ni-$d_{x^2-y^2}$ orbital character for this band.

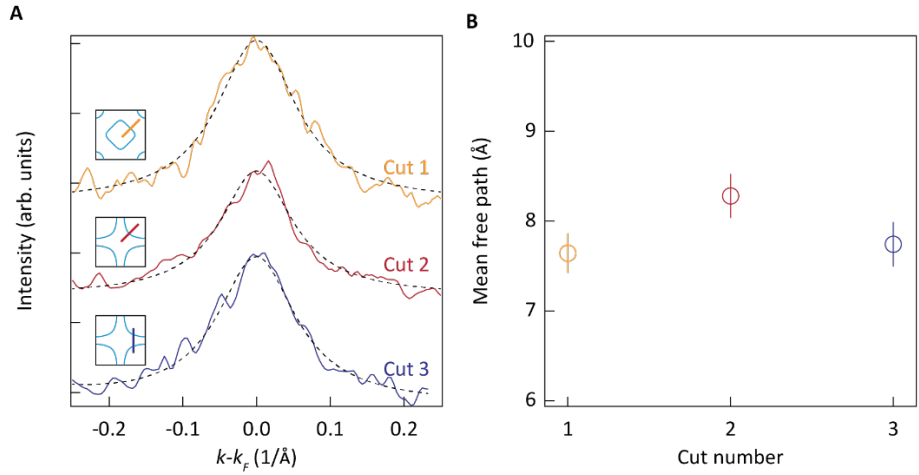

**Fig. S5.**
**Mean free paths of the Ni-$d_{x^2-y^2}$ derived bands. (A)** Representative linear-background-subtracted MDCs at the Fermi level for cuts shown in panel A, C, and E of Fig. 2 in the main text. Blank dashed lines are corresponding Lorentzian fittings. Inset shows the corresponding cuts in momentum space. **(B)** Corresponding inelastic mean free path calculated from the inverse number of the full width at half maximum (FWHM) of Lorentzian fits to these MDCs. Error bar comes from the fitting uncertainty.

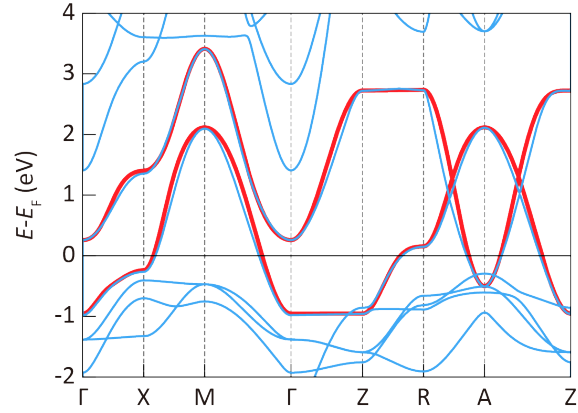

**Fig. S6.**
**Downfolding band structure for the low-energy effective model.** Comparison between the DFT-calculated band structure (blue lines) and the band structure computed from the low-energy effective model (red lines) shown in Eq. (1) in the supplementary materials.

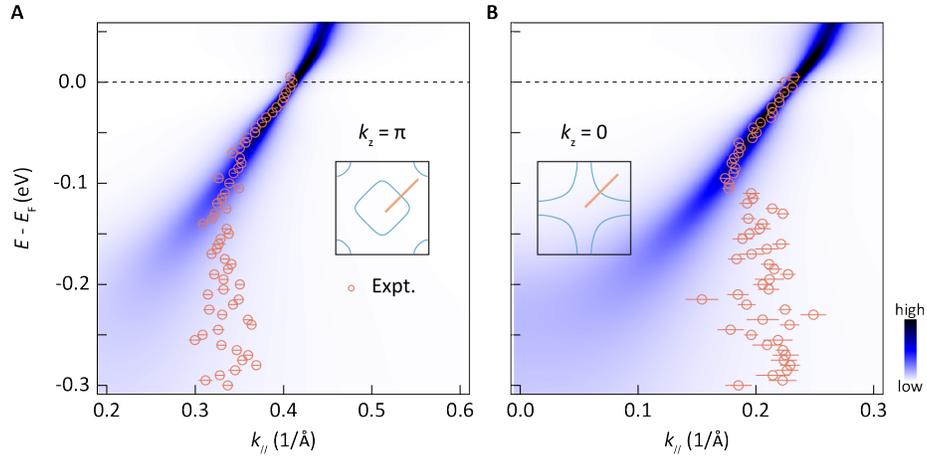

**Fig. S7.**
**Additional comparison between measured dispersion and DMFT results for Ni-$d_{x^2-y^2}$ derived bands. (A)** DMFT spectral function for measured ARPES spectra shown in Fig. 2A in the main text. **(B)** Same as (A) but corresponds to Fig. 2E in the main text.

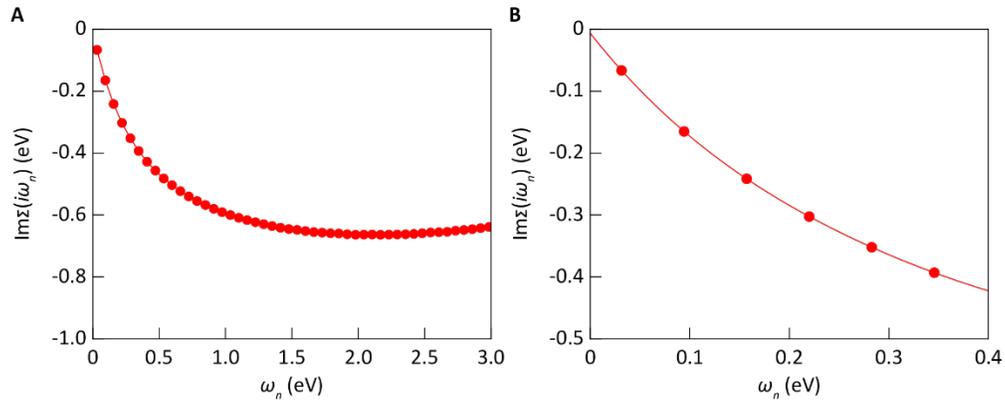

**Fig. S8.**
**The imaginary part of the self-energy of Ni-$d_{x^2-y^2}$ orbital in DMFT calculations. (A)** Frequency-dependent imaginary part of the self-energy with $U_d = 3.9$ eV. The red dots are from DMFT calculations, while the curves are the corresponding fourth-order polynomial fitting. **(B)** Enlarged view near zero frequency in (A).

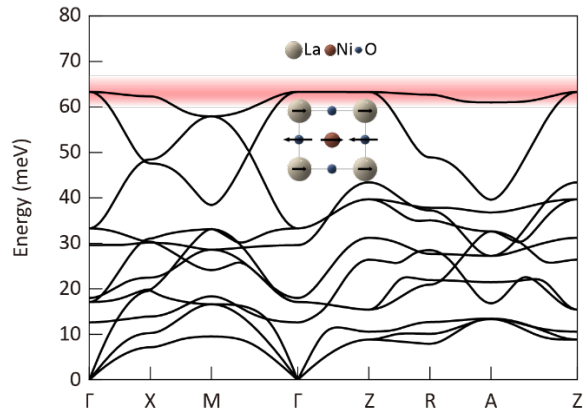

**Fig. S9.**
**Calculated phonon band structures of LaNiO$_2$.** The dispersion anomaly in ARPES spectra in Fig. 2 coincide with the highest phonon frequency at $\Gamma$ point, as shown by the red shaded region. Note that the Sr doping in real system will push the phonon frequency to a higher frequency due to lighter atomic mass of Sr than La. Inset shows the schematic drawing of atomic vibration mode associating with the highest phonon frequency, which is an in-plane full-breathing stretching mode as that in optimally-doped cuprates.

**Table S1.**

The onsite energy and hopping matrix elements in the *ds* model. All the units are eV.

| | | | |
|---|---|---|---|
| in $\epsilon_d(\mathbf{k})$ | $t_d^{100} = -0.384$ | $t_d^{110} = 0.053$ | $t_d^{200} = -0.057$ |
| in $\epsilon_s(\mathbf{k})$ | $E_{ds} = 1.077$ | $t_s^{100} = 0.004$ | $t_s^{110} = -0.050$ |
| | $t_s^{101} = -0.199$ | $t_s^{111} = 0.077$ | $t_s^{001} = -0.126$ |
| in $V_{ds}(\mathbf{k})$ | $t_{ds}^{100} = 0.088$ | $t_{ds}^{200} = 0.028$ | |